\documentclass[aps,prd,showpacs,superscriptaddress,twocolumn]{revtex4}
\usepackage{amsmath, amssymb}
\usepackage{graphicx}
\usepackage{dcolumn}
\usepackage{bm}
\newcommand{\wick}[1]{:\!{#1}\!:}

\begin{document}


\title{The $\Lambda$CDM-model in quantum field theory on curved spacetime and Dark Radiation}



\author{Thomas-Paul Hack}\email{hack@dima.unige.it}

\affiliation{Dipartimento di Matematica, Universit\`a degli Studi di Genova, I-16146 Genova, Italy.}


\date{\today}

\begin{abstract}
In the standard model of cosmology, the universe is described by a Robertson-Walker spacetime, while its matter/energy content is modeled by a perfect fluid with three components corresponding to matter/dust, radiation, and a cosmological constant. On the other hand, in particle physics matter and radiation are described in terms of quantum field theory on Minkowski spacetime. We unify these seemingly different theoretical frameworks by analysing the standard model of cosmology from first principles within quantum field theory on curved spacetime: assuming that the universe is homogeneous and isotropic on large scales, we specify a class of quantum states whose expectation value of the energy density is qualitatively and quantitatively of the standard perfect fluid form up to potential corrections. Qualitatively, these corrections depend on new parameters not present in the standard $\Lambda$CDM-model and can account for e.g. the phenomenon of Dark Radiation ($N_\text{eff}>3.046$), having a characteristic signature which clearly deviates from other potential Dark Radiation sources such as e.g. sterile neutrinos. Quantitatively, we find that our more fundamental model can be perfectly matched to observational data, such that we arrive at a natural and fundamental extension of the $\Lambda$CDM-model.
\end{abstract}

\keywords{keywords to be decided}
\pacs{04.62.+v, 98.80.-k, 95.36.+x, 95.35.+d}

\maketitle

\section{Introduction}
\label{sec_intro} 

Quantum field theory on curved spacetime (QFT on CST, see e.g. the monographs and reviews \cite{Bar:2009zzb, Benini:2013fia, Birrell:1982ix, Hack:2010iw, Mukhanov:2007zz, Parker:2009uva, Wald:1995yp}) is a framework in which matter is modelled as quantum fields propagating on a classical curved spacetime which is treated according to the principles of General Relativity. As such, QFT on CST is the simplest and natural generalisation of quantum field theory on Minkowski spacetime which takes into account that only in certain regimes a flat Minkowski spacetime is a good  description for our universe. Taking the spacetime as being classical, QFT on CST is presumably only valid in situations where the spacetime curvature scale is below the Planck scale and thus QFT on CST is the ideal framework for describing physics in the regime of medium-sized spacetime curvature, e.g. in the vicinity of black holes or in the early universe.

Indeed, quantum field theory on curved spacetime is an important ingredient in modern theoretical cosmology, as quantum fluctuations of the scalar field(s) responsible for the rapid expansion of the universe in the scenario of Inflation are believed to be the seeds of the structures in our present universe, see e.g. \cite{Mukhanov:2005sc, Straumann:2005mz} and the recent works \cite{Eltzner:2013soa, Pinamonti:2013zba}, and the analysis of these fluctuations is done in the framework of QFT on CST. However, other aspects of theoretical cosmology are usually not dealt with within QFT on CST, but simplified and less fundamental descriptions are used.

According to the Standard Model of Cosmology -- the $\Lambda$CDM-model -- our universe contains matter, radiation, and Dark Energy, whose combined energy density determines the expansion of the universe, see for example \cite{Ellis, Mukhanov:2005sc}. In the $\Lambda$CDM-model, these three kinds of matter-energy are modelled macroscopically as a perfect fluid and are thus completely determined by an energy density $\rho$ and a pressure $p$, with different equations of state $p = p(\rho) = w \rho$, $w=0, \frac13, -1$ for matter, radiation and Dark Energy (assuming that the latter is just due to a cosmological constant) respectively. Indeed, the terms "matter" and "radiation" in the context of cosmology subsume all matter-energy with the respective macroscopic equation of state such that e.g. "radiation" does not mean only electromagnetic radiation, but also the three left-handed neutrinos present in Standard Model of Particle Physics (SM) and possibly so-called Dark Radiation,  and "matter" means both the baryonic matter which is well-understood in the SM and Dark Matter. Here, Dark Matter and Dark Radiation both quantify contributions to the macroscopic matter and radiation energy densities which exceed the ones expected from our knowledge of the SM and are believed to originate either from particles respectively fields not present in the SM or from geometric effects, i.e. modifications of General Relativity. 

Notwithstanding, at least the contributions to the macroscopic matter and radiation energy densities which are in principle well-understood originate microscopically from particle physics, thence it should be possible to derive those from first principles within QFT on CST. However, in the standard literature usually a mixed classical/quantum analysis is performed on the basis of effective Boltzmann equations in which the collision terms are computed within QFT on flat spacetime whereas the expansion / curvature of spacetime is taken into account by means of redshift/dilution-terms, see e.g. \cite{Kolb:1990vq}. After a sufficient amount of cosmological expansion, i.e. in the late universe, the collision terms become negligible and the energy densities of matter and radiation just redshift as dictated by their equation of state.

In this work we aim to improve on this situation and to demonstrate that it is indeed possible to understand the $\Lambda$CDM-model from first principles within quantum field theory on curved spacetime: we model matter and radiation by quantum fields propagating on a cosmological spacetime and we show that there exist states for these quantum fields in which the energy density has the form assumed in the $\Lambda$CDM-model up to small corrections. Indeed, we find that these small corrections are a possible explanation for the phenomenon of Dark Radiation, which shows that a fundamental analysis of the $\Lambda$CDM-model is not only interesting from the conceptual point of view but also from the phenomenological one.

Due to the complexity of the problem and for the sake of clarity we shall make some simplifying assumptions. On the one hand, we shall model both matter and radiation by scalar and neutral quantum fields for the ease of presentation, but all concepts and principal constructions we shall use have been developed for fields of higher spin and non-trivial charge as well and we shall mention the relevant literature whenever appropriate. Thus, a treatment taking into account these more realistic fields is straightforward. On the other hand, we shall consider only non-interacting quantum fields and thus the effects of the field interactions which presumably played an important role in the early universe will only appear indirectly as characteristics of the states of the free quantum fields in our description. Notwithstanding, we stress that all concepts necessary to extend our treatment to interacting fields are already developed and we shall point out suitable references in due course. Finally, in this work we are only interested in modelling the history of the universe from the time of Big Bang Nucleosynthesis (BBN) until today, thus we shall in particular not be concerned with the era of Inflation and possible modifications thereof. Note that the restriction to the post-BBN era also justifies our approximation of considering non-interacting quantum fields, as one usually assumes that field interactions can be neglected on cosmological scales after electron-positron annihilation, which happened roughly at the same time as BBN \cite{Kolb:1990vq}. 

Our paper is organised as follows: in Section \ref{sec_quant} we review the quantization of the Klein-Gordon field on general curved spacetimes as well as the important concept of Hadamard states, which are the suitable generalisations of vacuum states and excitations thereof from Minkowski spacetime to curved spacetimes. As our aim is to make this work as accessible as possible to non-experts, we shall omit most technical details in this section. In the subsequent Section \ref{sec_semicla} we introduce the semiclassical Einstein equation which describes the influence of the quantum fields on the background spacetime and discuss related conceptual issues. Then we confine ourselves to cosmological spacetimes in Section \ref{sec_thermal} and review certain generalisations of thermal states on Minkowski spacetime to cosmological spacetimes which are suitable for cosmology. In Section \ref{sec_energy} we compute the energy density in these states and show that it matches the energy density in the $\Lambda$CDM-model up to small corrections. We analyse these corrections in more detail in Section \ref{sec_dark} and find that they can provide a natural explanation for Dark Radiation. Afterwards we conclude the paper in Section \ref{sec_conclusions}.

\section{Quantization of the Klein-Gordon field on general curved spacetimes}
\label{sec_quant} 

We first review the quantization of a free neutral scalar field $\phi$ on a general curved spacetime before analysing the case of cosmological spacetimes in more detail. Here, a curved spacetime is a tuple $(M,g)$, where $M$ is a four-dimensional smooth (i.e. infinitely often differentiable) manifold and $g$ is a smooth metric on $M$ with signature $(+,-,-,-)$; we shall often abbreviate $(M,g)$ by $M$ for simplicity. The classical neutral scalar field is characterised by satisfying the \emph{Klein-Gordon equation}
\begin{equation}\label{eq_KG}
P\phi:=\left(\Box + \xi R + m^2\right)\phi =0,
\end{equation}
$$\Box:=\nabla^\mu \nabla_\mu\,,$$
where $\nabla$ is the Levi-Civita covariant derivative associated to $g$, $R$ is the corresponding Ricci curvature scalar, $\xi$ denotes the coupling of $\phi$ to $R$ and $m$ is the mass of the scalar field. For technical reasons one often assumes that the spacetime $(M,g)$ is \emph{globally hyperbolic} in order to guarantee that \eqref{eq_KG} has unique solutions for any initial conditions given on an equal-time surface, see e.g. \cite{Bar:2009zzb, Hack:2010iw} for details; the Robertson-Walker spacetimes used in cosmology belong to this class.

In Minkowski spacetime, (time) translation invariance is an important ingredient in the usual construction of QFTs: first of all it is necessary in order to have a well-defined notion of \emph{energy}; given this, one can define a \emph{vacuum state} $\Omega_0$ and the associated notions of \emph{creation/annihilation operators}, \emph{particles} and \emph{Fock space} by requiring that all excitations of the vacuum have positive energy. General curved spacetimes, e.g. cosmological spacetimes, are not invariant under time-translations and thus a unique notion of vacuum does not exist. One may be able to find several candidates for generalised vacuum states, as we will also discuss in the following, but choosing any two of them, say $\Omega_A$ and $\Omega_B$, one typically finds because of the infinitely many degrees of freedom of a quantum field that $\Omega_A$ contains infinitely many particles w.r.t. $\Omega_B$, s.t. the two states are not unitarily equivalent. In order to discuss the quantization of the Klein-Gordon on arbitrary curved spacetimes it thus seems advisable to consider a more fundamental approach to quantization, the so-called \emph{algebraic approach} \cite{Haag:1992hx}. In this framework, one first defines an abstract algebra ${\cal A}$ which encodes all fundamental algebraic relations of the quantum fields, e.g. equations of motion and canonical (anti)commutation relations, and then analyses the possible states, i.e. the possible representations of ${\cal A}$ as operators on a Hilbert space. 

In the case of the neutral scalar field, the algebra ${\cal A}_{KG}$ encodes the Klein-Gordon equation \eqref{eq_KG}, the reality condition $\phi(x)^*=\phi(x)$ and the canonical commutation relations ($\hbar=1$)
\begin{equation}
\label{eq_CCR}
\left[\phi(x),\phi(y)\right]=i G(x,y)
\end{equation}
where $G(x,y)=G_-(x,y)-G_+(x,y)$ is the \emph{Pauli-Jordan commutator function} constructed from the advanced/retarded Green's functions $G_\pm(x,y)$ which satisfy 
$$P_xG_\pm(x,y)=\delta(x,y)$$
and vanish for $x$ in the past/future of $y$. This \emph{covariant} form of the canonical commutation relations is equivalent to the often used form where commutation relations of the field and its canonically conjugate momentum are specified at equal times, see e.g. \cite{Wald:1995yp, Bar:2009zzb, Hack:2010iw} for this and for a more detailed account of the quantization of the Klein-Gordon field on a curved spacetime in the algebraic framework.

As already anticipated, many unitarily-inequivalent Hilbert space representations of ${\cal A}_{KG}$ exist and, even if in general none of them is preferred, it would be nice to have criteria in order to select those which are of physical relevance. As the vacuum state $\Omega_0$ in Minkowski spacetime is pure and Gaussian (quasi-free), it seems advisable to concentrate on these states as well in curved spacetimes; they are completely characterised in terms of their two-point correlation function
$$\Omega(x,y):=\langle \phi(x) \phi(y) \rangle_\Omega$$
in particular. Note that the algebraic approach to QFT on CST is certainly powerful enough to treat non-Gaussian states with ease, we just concentrate on the Gaussian ones here in order to initially maintain as many properties of $\Omega_0$ as possible.
To arrive at further physical constraints on $\Omega$, we consider the functional properties of the vacuum two-point function $\Omega_0(x,y)$ in Minkowski spacetime. Indeed $\Omega_0(x,y)$ is divergent if $x$ and $y$ are connected by a light-like geodesic and the exact form of this singularity is important for assuring that all expressions appearing in \emph{Wick's theorem} for the product of normal-ordered quantities are well-defined. Using the seminal results in \cite{Radzikowski:1996pa, Radzikowski:1996ei}, the authors of \cite{Brunetti:1995rf,Brunetti:1999jn} showed that the correct generalisation of this singular structure to curved spacetime is the so-called \emph{Hadamard condition} \cite{Kay:1988mu}, which is satisfied by a state $\Omega$ if its two-point function is of the form
$$\Omega(x,y)= H(x,y)+W_\Omega(x,y)$$
$$H(x,y):=\frac{1}{8\pi^2}\left(\frac{U(x,y)}{\sigma(x,y)}+V(x,y)\log\left(\frac{\sigma(x,y)}{\lambda^2}\right)\right)$$
where $\sigma(x,y)$ is the \emph{half squared geodesic distance} between $x$ and $y$  \footnote{$x$ and $y$ must be ''sufficiently close'' in order to make sure that there exists a unique geodesic connecting them, which in turn guarantees that $\sigma(x,y)$ is well-defined. In the following we are only interested in the limit of $x\to y$ and thus we can safely assume that $H(x,y)$ is well-defined.}, $\lambda$ is an arbitrary length scale and $U(x,y)$, $V(x,y)$ and $W_\Omega(x,y)$ are smooth functions. $V(x,y)$ can be expanded w.r.t. to $\sigma(x,y)$, viz.
$$V(x,y)=\sum\limits_{n=0}^\infty V_n(x,y) \sigma^n(x,y)\,,$$
and $U(x,y)$ and all $V_n(x,y)$ can be completely specified in terms of (derivatives of) the metric $g$, and the parameters $\xi$ and $m$ appearing in \eqref{eq_KG} in a recursive manner. Thus, the \emph{Hadamard singularity} $H(x,y)$ is universal among all Hadamard states $\Omega$ and only the regular part of their two-point function $W_\Omega(x,y)$ depends on $\Omega$. In general it is not possible to prove the convergence of the series $V(x,y)$, but while this is dissatisfactory from the the conceptual point of view, it is not relevant for the computation performed in the following, as there in the limit $x\to y$ all but a finite number of terms in the series vanish.

We have briefly reviewed the algebraic approach to quantization and the notion and physical and mathematical properties of Hadamard states only for the free neutral scalar field, but these concepts have been developed for fields of higher spin, gauge fields and interacting fields as well. For this and further details we refer the reader to e.g. \cite{Brunetti:1999jn, Sahlmann:2000zr, Fewster:2003ey, Hollands:2007zg, Sanders:2008kp, Dappiaggi:2009xj, Hack:2010iw, Fredenhagen:2011mq, Dappiaggi:2011cj, Fewster:2012bj, Hack:2012dm}.

\section{The semiclassical Einstein equation}
\label{sec_semicla} 

The starting point for the understanding of the $\Lambda$CDM-model within quantum field theory in curved spacetime is the \emph{semiclassical Einstein equation}
\begin{equation}
\label{eq_semicla}
G_{\mu\nu}=8\pi G \langle \wick{T_{\mu\nu}}\rangle_\Omega\,,
\end{equation}
where $G_{\mu\nu}$ is the Einstein tensor, $G$ is Newton's constant, $\wick{T_{\mu\nu}}$ is the regularised stress-energy tensor of all quantum fields in the model, and $\Omega$ is a suitable state. There are several conceptual issues related to this equation of which we would like to mention a few in the following. First of all in \eqref{eq_semicla} one equates a classical quantity with a probabilistic one, which makes sense only if the fluctuations of the latter are small. From the mathematical properties of Hadamard states it follows that the fluctuations of (a suitably regularised) $\wick{T_{\mu\nu}}$ are finite in any Hadamard state $\Omega$ \cite{Brunetti:1995rf}, whereas a discussion of the actual size of the fluctuations can be found in \cite{Pinamonti:2010is}. However, as already mentioned it is not at all clear how to pick a preferred state among the class of all Hadamard states, and in fact writing down the semiclassical Einstein equations implies that we are able to specify a map $\Omega(M)$ which assigns to each spacetime $M$ a Hadamard state in a coherent manner. Unfortunately, it has been explicitly proven in \cite{Fewster:2011pe} that this is impossible if one wants to do this for all (globally  hyperbolic) spacetimes. However, a possible way out is to restrict the allowed class of spacetimes. Indeed, as we will review in the next section, it is possible to assign coherently a Hadamard state to each Robertson-Walker spacetime.

The next conceptual issue we would like to mention is the actual \emph{definition} of $\langle \wick{T_{\mu\nu}}\rangle_\Omega$ or $\wick{T_{\mu\nu}}$ respectively. In usual particle physics experiments we always measure the \emph{difference} of the expectation value of $\wick{T_{\mu\nu}}$ in two states, e.g. the vacuum and a many-particle state. However, gravity is sensitive to the \emph{absolute} value of $\langle \wick{T_{\mu\nu}}\rangle_\Omega$, thus the unambiguous specification of $\langle \wick{T_{\mu\nu}}\rangle_\Omega$ corresponds to a specification of a ''zero point'' in the absolute energy scale, but this is impossible within quantum field theory in curved spacetime. In more detail, one could ask the question: what is the most general expression for $\wick{T_{\mu\nu}}$ which is compatible with all physical consistency conditions I can impose? The conditions one could impose are: a) correct commutation relations with other observables, b) covariant conservation $\nabla^\mu\wick{T_{\mu\nu}}=0$, c) $\wick{T_{\mu\nu}(x)}$ should be a \emph{local} object and depend only on $x$ in a suitable sense, but a state is non-local on account of the equations of motion; thus, the observable $\wick{T_{\mu\nu}(x)}$ should be defined in a state-independent manner. Indeed, one can show that there is no unique expression which satisfies all these conditions (and further technical ones) \cite{Walda, Mo03, HW04}. The most general expression for the expectation value $\langle \wick{T_{\mu\nu}}\rangle_\Omega$ turns out to be
\begin{equation}\label{eq_nonunique}
\langle\wick{T_{\mu\nu}}\rangle_\Omega = \langle \wick{T_{\mu\nu}}_0\rangle_\Omega + \alpha_1 g_{\mu\nu} + \alpha_2 G_{\mu\nu} + \alpha_3 I_{\mu\nu} + \alpha_4 J_{\mu\nu}\,.
\end{equation} 

Here, $\alpha$ and $\beta$ can be interpreted as a (renormalisation of) the cosmological constant and a renormalisation of Newton's constant, whereas $I_{\mu\nu}$ and $J_{\mu\nu}$ are conserved local curvature tensors which contain fourth derivatives of the metric \cite{Mo03, HW04, Wald3} and are obtained as functional derivatives with respect to the metric of the Lagrangeans $\sqrt{-g}R^2$ and $\sqrt{-g}R_{\mu\nu}R^{\mu\nu}$ respectively; these higher-derivative contributions are usually ruled out in classical General Relativity but one can show that they can not be avoided in QFT on CST \cite{Wald3}. Moreover, a ''model'' $\langle \wick{T_{\mu\nu}}_0\rangle_\Omega$ is \cite{Mo03}
\begin{equation}\label{eq_deftmunu}\langle \wick{T_{\mu\nu}}_0\rangle_\Omega:=\lim\limits_{x\to y} \left(D_{\mu\nu}-\frac13 g_{\mu\nu} P_x\right)\left(\Omega(x,y)-H(x,y)\right)
\end{equation}
\begin{align*}D_{\mu\nu}&:=(1-2\xi)g^{\nu^\prime}_\nu\nabla_{\mu}\nabla_{\nu^\prime}-2\xi\nabla_{\mu}\nabla_\nu-\xi G_{\mu\nu}\\&\quad+g_{\mu\nu}\left\{2\xi \square_x+\left(2\xi-\frac12\right)g^{\rho^\prime}_\rho\nabla^\rho\nabla_{\rho^\prime}+\frac12 m^2\right\}\,.
\end{align*} 
Here $g^{\nu^\prime}_\nu$ denotes the parallel transport of a vector from $x$ to $y$ along the geodesic connecting $x$ and $y$, the form of $D_{\mu\nu}$ follows directly from the classical stress-energy tensor of the scalar field, viz.
\begin{align*}T_{\mu\nu}&=(1-2\xi)\left(\nabla_\nu \phi\right) \nabla_{\mu}\phi-2\xi\phi\nabla_{\mu}\nabla_\nu\phi-\xi G_{\mu\nu}\phi^2\\&\quad+g_{\mu\nu}\left\{2\xi \phi\square\phi+\left(2\xi-\frac12\right)\left(\nabla^\rho\phi\right)\nabla_{\rho}\phi+\frac12 m^2\phi^2\right\}\end{align*}
and the modification term $-\frac13 g_{\mu\nu} P_x$ is necessary in order to have a covariantly conserved $\langle \wick{T_{\mu\nu}}_0\rangle_\Omega$ \cite{Mo03}. Note that $\langle \wick{T_{\mu\nu}}_0\rangle_\Omega$ is not unambiguously defined itself as it depends on the length scale $\lambda$ in $H(x,y)$. Indeed, if one changes the length scale $\lambda$ in $H(x,y)$ as appearing in \eqref{eq_deftmunu} to a new value $\lambda^\prime$, then $\langle \wick{T_{\mu\nu}}_0\rangle_\Omega$ changes by

\begin{equation}\label{eq_lchange}
{2\log\lambda/\lambda^\prime}{8\pi^2}\left(\frac{(6\xi-1)m^2G_{\mu\nu}}{12}-\right.\end{equation}$$\left. -
\frac{m^4g_{\mu\nu}}{8} +\frac{I_{\mu\nu}-3J_{\mu\nu}}{360}-\frac{(6\xi-1)^2I_{\mu\nu}}{144}\right)\,.$$

The parameters $\alpha_i$ are free parameters of the theory which are independent of the field content and the spacetime $M$ and can in principle be fixed by experiment, just like the mass $m$. In the following we will take the point of view that $\alpha_2$ is not a free parameter because Newton's constant has been measured already. In order to do this, we have to fix a value for the length scale $\lambda$ in the Hadamard singularity $H(x,y)$, we do this by confining $\lambda$ to be a scale in the range in which the strength of gravity has been measured. Because of the smallness of the Planck length, the actual value of $\lambda$ in this range does not matter as changing $\lambda$ in this interval gives a negligible contribution to $\langle \wick{T_{\mu\nu}}\rangle_\Omega$. Moreover, in the case of conformal coupling $\xi=\frac16$, which we shall assume most of the time, $\alpha_2$ is independent of $\lambda$ as one can infer from \eqref{eq_lchange}. One could also take a more conservative point of view and consider $\alpha_2$ to be a free parameter, in this case comparison with cosmological data, e.g. from Big Bang Nucleosynthesis, would presumably constrain $\alpha_2$ to be very small once $\lambda$ is in the discussed range.

Thus, we are left with three free parameters in the definition of $\langle \wick{T_{\mu\nu}}\rangle_\Omega$: one of them corresponds to the cosmological constant which is already a free parameter in classical General Relativity, whereas the other two parameters do not appear there and thus will by themselves  lead to an extension of the $\Lambda$CDM-model.

To close this section, we would like to highlight the point of view on the so-called {\it cosmological constant problem} taken in this work, as well as in most works on QFT on CST in the algebraic approach and e.g. the review \cite{Bianchi:2010uw}. It is often said that QFT {\it predicts} a value for the cosmological constant which is way too large in comparison to the one measured. This conclusion is reached by computing one or several contributions to the vacuum energy in Minkowski spacetime $\Lambda_\text{vac}$ and finding them all to be too large, such that, at best, a fine-tuned subtraction in terms of a negative bare cosmological constant $\Lambda_\text{bare}$ is necessary in order to obtain the small value $\Lambda_\text{vac}+\Lambda_\text{bare}$ we observe. In this work, we assume as already mentioned the point of view that it is not possible to provide an {\it absolute} definition of energy density within QFT on CST, and thus neither $\Lambda_\text{vac}$ nor $\Lambda_\text{bare}$ have any physical meaning by themselves; only $\Lambda_\text{vac}+\Lambda_\text{bare}$ is physical and measureable and any cancellation which happens in this sum is purely mathematical. The fact that the magnitude of $\Lambda_\text{vac}$ depends on the way it is computed, e.g. the loop or perturbation order, cf. e.g. \cite{Sola:2013gha}, is considered to be unnatural following the usual intuition from QFT on flat spacetime. However, it seems more convincing to us to accept that $\Lambda_\text{vac}$ and $\Lambda_\text{bare}$ have no relevance on their own, which does not lead to any contradiction between theory and observations, rather than the opposite.

In the recent work \cite{Holland:2013xya} it is argued that a partial and unambiguous relevance can be attributed to $\Lambda_\text{vac}$ by demanding $\Lambda_\text{bare}$ to be analytic in all coupling constants and masses of the theory; taking this point of view, one could give the contribution to $\Lambda_\text{vac}$ which is non-analytic in these constants an unambiguous meaning. Indeed the authors of \cite{Holland:2013xya} compute a non-perturbative and hence non-analytic contribution to $\Lambda_\text{vac}$, which turns out to be small. In the view of this, one could reformulate our statement in the above paragraph and say that contributions to $\Lambda_\text{vac}$ and $\Lambda_\text{bare}$ which are analytic in masses and coupling constants have no physical relevance on their own.

\section{States of interest on cosmological spacetimes}
\label{sec_thermal} 

After the review of the quantum theory of a free scalar field on general curved spacetimes we shall consider only cosmological spacetimes in the following. To wit, we assume that the spacetime is given by a spatially flat Robertson-Walker spacetime, i.e. a subset of $\mathbb{R}^4$ with the metric
$$ds^2 = dt^2 - a(t)^2 d\vec{x}^2\,.$$
The translational and rotational invariance of this metric in the spatial coordinates reflects the paradigm that our universe is homogeneous and isotropic on large scales, while we have chosen a Robertson-Walker spacetime without spatial curvature in order to simplify computations and because observations are compatible with the assumptions of vanishing spatial curvature \cite{Ade:2013zuv}.

As usual, $t$ is cosmological or co-moving time, whereas $a(t)$ is the {\it scale factor} whose expansion rate is the {\it Hubble rate}
$H:=\dot{a}/a$. Further possible time variables are the conformal time $\tau$, and, if $H$ is strictly positive (or negative), the scale factor $a$ itself as well as the {\it redshift} $z:=a_0/a-1$, where $a_0$ is the scale factor of today, usually set to $a_0=1$ by convention. These time variables are related by
$$dt = a d\tau = \frac{da}{aH} = -\frac{dz}{(1+z)H}\,.$$
In the following we shall always assume $H>0$ and change freely between these time variables; derivatives with respect to $t$ ($\tau$) shall be denoted by $\dot{\phantom{a}}$ (${\phantom{a}}^\prime$). Using $a$ or $z$ as time variables is often convenient because it does not require the explicit knowledge of $a$ as a function of $t$ or $\tau$. Moreover, the redshift $z$ is a direct observable in contrast to the other time parameters.

The goal of this section is to introduce a class of states which we believe to be a good model for the actual quantum states that describe the content of our universe on large scales, as will be justified in the next section. As most of the energy density in the $\Lambda$CDM-model is believed to be of thermal origin, we seek states which can be considered as generalised thermal states \footnote{The definition of strict thermal equilibrium states is only possible under the assumption of time-translation invariance, which is violated in cosmology since $H>0$.}. However, for this we first need to specify a good class of generalised vacuum states, which is what we shall do now. To this avail, we recall that a {\it pure} and {\it Gaussian} isotropic and homogeneous state for the Klein-Gordon field on a spatially flat Robertson-Walker spacetime is determined by a two-point correlation function of the form \cite{Lueders:1990np}
$$\Omega(x,y)=\frac{1}{8\pi^3 a(\tau_x)a(\tau_y)}\int\limits_{\mathbb{R}^3} d{\vec k}\, \overline{\chi_k(\tau_x)}\chi_k(\tau_y)e^{i\vec{k}(\vec{x}-\vec{y})}\,,$$
where the modes $\chi_k$ satisfy the ordinary differential equation
\begin{equation}\label{eq_modesode}
\left(\partial^2_\tau+k^2+m^2a^2 + \left(\xi-\frac16\right)R a^2\right)\chi_k(\tau)=0
\end{equation}
and the normalisation condition
\begin{equation}\label{eq_modesnormal}
\overline{\chi_k}^\prime{\chi_k}-\overline{\chi_k}{\chi^\prime_k}=i\,.
\end{equation}
Here, $k:= |\vec{k}|$ and $\overline{\cdot}$ denotes complex conjugation. Choosing a solution of the above equation for each $k$ amounts to  specifying the state. 

Distinguished states are the {\it adiabatic states} introduced in \cite{Parker:1969au}. They are specified by modes of the form
\begin{equation}\label{eq_adiabatic}\chi_k(\tau)=\frac{1}{\sqrt{\omega(k, \tau)}}\exp\left({-i \int^\tau_{\tau_0}\omega(k,\tau^\prime) d\tau^\prime}\right),\end{equation}
where $\omega(k,\tau)$ solves a non-linear differential equation in $\tau$ obtained by inserting this ansatz into \eqref{eq_modesode} and finding
$$\omega(k,\tau)^2=f(\omega(k,\tau)^{\prime\prime}, \omega(k,\tau)^\prime, \omega(k,\tau), a(\tau))$$
for a suitable function $f$. While this ansatz in principle holds for any state, the adiabatic states are specified by solving the differential equation for $\omega(k,\tau)$ iteratively as
$$\omega_{n+1}(k,\tau)^2:=f(\omega_n(k,\tau)^{\prime\prime}, \omega_n(k,\tau)^\prime, \omega_n(k,\tau), a(\tau))$$
starting from $\omega_0(k,\tau)=\sqrt{k^2+m^2a^2 + \left(\xi-\frac16\right)R a^2}$. Truncating this iteration after $n$ steps defines the adiabatic states of order $n$. Note that, while the resulting modes satisfy the normalisation condition \eqref{eq_modesnormal} exactly, they satisfy \eqref{eq_modesode} only up to terms which vanish in the limit of constant $a$ or of infinite $k$ and/or $m$. Thus they constitute only approximate states. This can be cured by using the adiabatic modes of order $n$ only for the specification of the initial conditions for exact solutions of \eqref{eq_modesode}, see  \cite{Lueders:1990np}. Regarding the UV properties of such defined `proper' adiabatic states, it has been shown in \cite{Junker} (for spacetimes with compact spatial sections) that they are in general not as UV-regular as Hadamard states, but that they approach the UV-regularity of Hadamard states in a certain sense in the limit of large $n$. In the following we shall often use the `improper' adiabatic modes of order $0$, $\chi_{0,k}(\tau):=\exp(-i\int^\tau_{\tau_0}\omega_0(k,\tau^\prime) d\tau^\prime))/\sqrt{2 \omega_0(k,\tau)}$. Adiabatic states have also been constructed for Dirac fields, see \cite{Hollands:1999fc, Landete:2013axa}, and general curved spacetimes \cite{Hollands:1999fc, Junker}.

A further class of states of interest in cosmology, and in fact our candidates for generalised vacuum states, are the {\it states of low energy} (SLE) introduced in \cite{Olbermann:2007gn}, motivated by results of \cite{Fewster:1999gj}. These states are defined by minimising the energy density per mode $\rho_k$
$$\rho_k:=\frac{1}{16 a^4 \pi^3}\left(|\chi_k^\prime|^2+\left(\xi -\frac16\right)2a\Re{\chi_k^\prime\overline{\chi_k}}+\right.$$$$\left.+\left(k^2+m^2a^2-\left(\xi -\frac16\right)H^2a^2\right)|\chi_k|^2\right)$$
integrated in (cosmological) time with a sampling function $f$ and thus loosely speaking minimise the energy in the time interval where the sampling function is supported. The minimisation is performed by choosing arbitrary basis modes $\chi_k$ and then determining the Bogoliubov coefficients $\lambda(k)$, $\mu(k)$ with respect to these modes, such that the resulting modes of the state of low energy are 
$$\chi_{f,k}=\lambda(k) \chi_k+\mu(k) \overline{\chi_k}$$
with
$$\lambda(k):=e^{i(\pi -\arg c_2(k))}\sqrt{\frac{c_1(k)}{\sqrt{c_1(k)^2-|c_2(k)|^2}}+\frac12}$$
$$\mu(k) := \sqrt{|\lambda(k)|^2-1}$$
$$c_1(k):= \frac{1}{2}\int dt f(t) \frac{1}{a^4}\left(|\chi_k^\prime|^2+\left(\xi -\frac16\right)2a\Re{\chi_k^\prime\overline{\chi_k}}+\right.$$$$\left.+\left(k^2+m^2a^2-\left(\xi -\frac16\right)H^2a^2\right)|\chi_k|^2\right)$$
$$c_2(k):= \frac{1}{2}\int dt f(t) \frac{1}{a^4}\left({\chi_k^\prime}^2+\left(\xi -\frac16\right)2a\chi_k^\prime\chi_k+\right.$$$$\left.+\left(k^2+m^2a^2-\left(\xi -\frac16\right)H^2a^2\right)\chi_k^2\right)\,.$$
\cite{Olbermann:2007gn} only discusses the case of minimal coupling, i.e. $\xi=0$ and proves that the corresponding SLE satisfy the Hadamard condition for sampling functions $f$ which are smooth and of compact support in time. However, we shall use these states for the case of conformal coupling $\xi=\frac16$, and, although we do not prove that they satisfy the Hadamard condition, we shall find them to be at least regular enough for computing the energy density. Moreover, it is not difficult to see that the SLE construction yields the {\it conformal vacuum} 
$$\chi_{f,k}(\tau)=\frac{1}{\sqrt{2k}}e^{-ik\tau}\,,$$
and thus a Hadamard state \cite{Pinamonti:2008cx}, for all sampling functions $f$ in the massless case. This demonstrates both that the SLE construction for $\xi=\frac16$ yields Hadamard states at least in special cases and that states of low energy deserve to be considered as generalised vacuum states on curved spacetimes. The SLE construction has recently been generalised to spacetimes with less symmetry in \cite{Them:2013uka}.

A conceptual advantage of states of low energy is the fact that they can be consistently defined an all Robertson-Walker spacetimes at once just by specifying the sampling function $f$ once and for all (with respect to e.g. cosmological time and a fixed origin of the time axis). Thus, they solve the conceptual problem mentioned in Section \ref{sec_semicla}, namely the necessity to specify a state in way which does not depend on the spacetime in order for the semiclassical Einstein equation to be well-defined a priori.

We now proceed to construct the anticipated generalised thermal states on the basis of states of low energy. To this avail, we recall a result of \cite{Dappiaggi:2010gt}: given a pure, isotropic and homogeneous state, i.e. a set of modes $\chi_k$, one can construct generalised thermal states with a two-point correlation function of the form

$$\Omega(x,y)=\frac{1}{8\pi^3 a(\tau_x)a(\tau_y)}\int\limits_{\mathbb{R}^3} d{\vec k}\,,e^{i\vec{k}(\vec{x}-\vec{y})}\times$$$$\times \frac{\overline{\chi_k(\tau_x)}\chi_k(\tau_y)}{1-e^{-\beta k_0}}+\frac{\chi_k(\tau_x)\overline{\chi_k(\tau_y)}}{e^{\beta k_0}-1}\,,$$
with
$$k_0:=\sqrt{k^2+m^2a^2_F}\,.$$
It has been shown in \cite{Dappiaggi:2010gt} that for the case of conformal coupling, special Robertson-Walker spacetimes and particular generalised vacuum modes $\chi_k$ on these spacetimes, these states satisfy certain generalised thermodynamic laws and the Hadamard condition, and one can show that they satisfy the Hadamard condition on general Robertson-Walker spacetimes if the pure state specified by $\chi_k$ is already a Hadamard state by using (slight generalisations of) results of \cite{Pinamonti:2010is}.

We shall assume in the following that the quantum fields in our model are in a generalised thermal state of the form as above, with generalised vacuum modes $\chi_k$ specified by a state of low energy with suitable sampling function $f$. If $m>0$, the phenomenological interpretation of these states is that they are the quantum state of a massive field which has been in thermal equilibrium in the hot early universe and has departed from this equilibrium at the `freeze-out time' $a=a_F$. In the massless case, these states are just conformal rescalings of the thermal equilibrium state with temperature $1/\beta$ in Minkowski spacetime. 

The generalised thermal states we use here have been discussed also for Dirac fields, see \cite{Dappiaggi:2010gt}. Moreover, we would like to mention that several definitions of generalised thermal states on curved spacetimes have been proposed so far, including {\it almost equilibrium states} \cite{Kusku:2008zz} and {\it local thermal equilibrium states} \cite{Schlemmer, VerchRegensburg}. A comparison of these different proposals in the context of cosmological applications would certainly be interesting, but is beyond the scope of this work.

\section{Computation of the energy density}
\label{sec_energy} 

We now approach the first main result of this work, a demonstration that the energy density in the $\Lambda$CDM-model can be reproduced from first principles within quantum field theory in curved spacetime. To this avail, we consider the following setup: we model radiation by a conformally coupled massless scalar quantum field, and matter/dust by a conformally coupled massive scalar quantum field. We choose conformal coupling also for the massive scalar field because this simplifies analytical computations a lot and we also found numerical computations to be more stable with this value of non-minimal coupling to the curvature.  Moreover, both quantum fields are assumed to be in generalised thermal equilibrium states as introduced in the previous section, where the state and field parameters $\beta$ (possibly different values for the two quantum fields), $m$ and $a_F$, as well as the sampling functions $f$ determining the generalised vacuum states of the two fields, are considered to be undetermined for the time being. Let us stress once more that there is no principal obstruction for formulating this model with more realistic quantum fields of higher spin, we just consider scalar quantum fields for simplicity and ease of presentation.

An exact computation of the energy density of the two quantum fields in the generalised thermal states would require to solve the coupled system -- the so-called {\it backreaction problem} -- consisting of the quantum fields propagating on a Robertson-Walker spacetime, which in turn is a solution of the {\it semiclassical Friedmann equation}
\begin{equation}
H^2 = \frac{8 \pi G}{3} \left(\rho^0 + \rho^m\right)\,,
\end{equation}
where $\rho^m=\langle :T^m_{00}:\rangle_{\Omega^m}$, $\rho^0=\langle :T^0_{00}:\rangle_{\Omega^0}$ are the energy densities of the two quantum fields in the respective generalised thermal states and the $00$-component of the stress-energy tensor is here taken with respect to cosmological time $t$. An exact solution of the backreaction problem is quite involved, as it requires solving simultaneously the mode equation \eqref{eq_modesode} for all $k$ and the semiclassical Friedmann equation. Notwithstanding there have been quantitative numerical treatments of the backreaction problem, see e.g. \cite{Anderson1, Anderson:1984jf, Anderson:1985cw, Anderson:1985ds, Baacke:2010bm}, as well as numerous qualitative treatments including \cite{Eltzner:2010nx}, where the backreaction problem in Robertson-Walker spacetimes is set up in full generality from the point of view of the algebraic approach to QFT on CST, \cite{DFP}, where the same point of view is considered and the coupled system is solved exactly for conformally coupled massless scalar quantum fields and approximately for massive ones, and \cite{Pinamonti:2008cx}, where a variant of the backreaction problem is solved exactly for conformally coupled massive scalar quantum fields in the vicinity of the Big Bang on Robertson-Walker spacetimes with a lightlike Big Bang hypersurface.

However, in this work we follow a simplified strategy in order to avoid solving the full backreaction problem, which is justified in view of our aim. We assume that the two quantum fields in our model are propagating on a Robertson-Walker spacetime which is an exact solution of the Friedmann equation in the $\Lambda$CDM-model, i.e.
\begin{equation}\label{eq_friedmannlcdm}
\frac{H^2}{H^2_0}=\frac{\rho_\text{$\Lambda$CDM}}{\rho_0}=\Omega_\Lambda + \frac{\Omega_m}{a^3} + \frac{\Omega_r}{a^4}\,,
\end{equation}
where $H_0\simeq 10^{-33}$eV denotes the Hubble rate of today, the so-called {\it Hubble constant}, $\rho_0\simeq 10^{-11}$eV$^4$ is the energy density of today and $\Omega_\Lambda$, $\Omega_m$ and $\Omega_r$ denote respectively the present-day fraction of the total energy density contributed by the cosmological constant, matter/dust and radiation. For definiteness we consider the sample values $\Omega_m=0.30$, $\Omega_r = 10^{-4}$, $\Omega_\Lambda=1-\Omega_m-\Omega_r$, rather than currently measured values from e.g. the Planck collaboration \cite{Ade:2013zuv}, because the exact values are not essential for our results. Given this background spacetime, we strive to prove that the field and state parameters of our model, as well as the SLE sampling functions, can be adjusted in such a way that the energy density of the quantum fields in our model matches the one in the $\Lambda$CDM-model up to negligible corrections for all redshifts $z\in[0,10^9]$, i.e.
$$\frac{\rho^0 + \rho^m}{\rho_0}\simeq\Omega_\Lambda + \frac{\Omega_m}{a^3} + \frac{\Omega_r}{a^4}=\frac{\rho_\text{$\Lambda$CDM}}{\rho_0}\,.$$
Once we succeed to obtain this result, we have clearly solved the full coupled system in an approximative sense to a good accuracy in particular.

In order to compute the quantum energy density $\rho^0 + \rho^m$, we start from \eqref{eq_nonunique} and 
\eqref{eq_deftmunu}. The former equation parametrises the freedom in defining the energy density as an observable, whereas the latter gives a possible ``model definition''. The renormalisation freedom for the energy density is readily computed as $g_{00}=1$, $G_{00}=3H^2$ and
\begin{equation}\label{eq_J00}J_{00}=\frac13 I_{00}=6 \dot{H}^2 - 12 \ddot{H}H- 36 \dot{H} H^2\,.
\end{equation}
In order to compute the energy density for each quantum field following from \eqref{eq_deftmunu}, one has to first subtract the Hadamard singularity from the two-point correlation function of the given state and then to apply a suitable bidifferential operator followed by taking the coinciding point limit. As the states we consider here are given as integrals over spatial momenta, it seems advisable to try to re-write the Hadamard singularity also in this form, in order to perform a mode-by-mode subtraction and momentum space integral afterwards. This is indeed possible, as elaborated in \cite{DFP, Pinamonti:2010is, Eltzner:2010nx, Schlemmer, Degner}. The details are quite involved, thus we omit them and present directly the result. To this avail we follow \cite{Degner}, where results of \cite{Schlemmer} are used. In \cite{Degner} only the minimally coupled case $\xi=0$ is discussed, but it is not difficult to generalise the results there to arbitrary $\xi$. 

Doing this, we find the following result for the total energy density of the massless and massive conformally coupled scalar fields in the generalised thermal states.
\begin{align}\label{eq_totaled}\frac{\rho^0 + \rho^m}{\rho_0}=&\frac{\rho^m_\text{gvac}+\rho^0_\text{gvac}+\rho^m_\text{gth}+\rho^0_\text{gth}}{\rho_0}\\&+\gamma \frac{H^4}{H^4_0}+\Omega_\Lambda+\delta\frac{H^2}{H^2_0}+\epsilon\frac{J_{00}}{H^4_0}\notag
\end{align}
$$\gamma:=\frac{8\pi G H^2_0}{360 \pi^2}\qquad \Omega_\Lambda=\frac{8\pi G \alpha_1}{3 H^2_0}$$
$$\delta:=\frac{8\pi G \alpha_2}{3 H^2_0}\qquad \epsilon:=\frac{8\pi GH^2_0}{3 }(3\alpha_3 + \alpha_4)\,.$$
Here $\Omega_\lambda$, $\delta$ and $\epsilon$ parametrise the freedom in the definition of the energy density as per \eqref{eq_nonunique}. The number of free parameters in this equation has been reduced to three, because $I_{\mu\nu}$ and $J_{\mu\nu}$ are proportional in Robertson-Walker spacetimes, cf. \eqref{eq_J00}. As already discussed in Section \ref{sec_semicla}, we omit the freedom parametrised by $\delta$ in the following, as it renormalises the Newton constant and we consider this to be already given as an external input. For now we will also neglect the contribution parametrised by $\epsilon$, as it turns out to be negligible for $0\le\epsilon\ll1$; we will analyse the influence of this new term, which does not appear in the $\Lambda$CDM-model, separately in the next section. Thus, for the remainder of this section,  $\Omega_\lambda$ parametrises the residual freedom in the definition of the quantum energy density. The term proportional to $\gamma$, which is also not present in  the $\Lambda$CDM-model, appears due to the so-called {\it trace anomaly}, which is a genuine quantum and moreover state-independent contribution to the quantum stress-energy tensor, see e.g. \cite{Wald3}. This term is fixed by the field content, i.e. by the number and spins of the fields in the model and always proportional to $H^4$, barring contributions proportional to $J_{00}$ which we prefer to subsume in the parameter $\epsilon$. We have given here the value of $\gamma$ for two scalar fields, see Table 1
on page 179 of \cite{Birrell:1982ix} for the values in case of higher spin. As $\gamma\simeq 10^{-122}$ and $H< H_0 z^2$ in the $\Lambda$CDM-model for large redshifts, this term can be safely neglected for $z<10^9$. Finally, the remaining terms in \eqref{eq_totaled} denote the genuinely quantum state dependent contributions to the energy densities of the two quantum fields. We have split these contributions into parts which are already present for infinite inverse temperature parameter $\beta$ in the generalised thermal states, and thus could be considered as contributions due to the generalised vacuum states ($\rho^m_\text{gvac}$, $\rho^0_\text{gvac}$), and into the remaining terms, which could be interpreted as purely thermal contributions ($\rho^m_\text{gth}$, $\rho^0_\text{gth}$). Note that $\rho^m_\text{gvac}$, $\rho^0_\text{gvac}$ are not uniquely defined in this way, but only up to the general renormalisation freedom of the quantum energy density, i.e. one could ``shuffle parts of'' $\Omega_\Lambda$, $\delta$ and $\epsilon$ into e.g. $\rho^m_\text{gth}$ and vice versa, without changing any physical interpretation of the total energy density. With this in mind, the state-dependent contributions read as follows, where the massless case is simply obtained by inserting $m=0$, and we give here the result for arbitrary coupling $\xi$ for completeness.

\begin{widetext}
\begin{gather}
\rho^m_{\text{gvac}} = \frac{1}{2\pi^3}\int\limits_0^\infty dk k^2\left\{\frac{1}{2a^4}\left(|\chi_k^\prime|^2+\left(\xi -\frac16\right)2a\Re{\chi_k^\prime\overline{\chi_k}}+\left(k^2+m^2a^2-\left(\xi -\frac16\right)H^2a^2\right)|\chi_k|^2\right)\right.\notag\\
\label{eq_rhofinal}\left.-\left(\frac{k}{2 a^4}+\frac{m^2-6H^2(\xi -\frac16)}{4 a^2 k}+\Theta(k-ma)\frac{-m^4 + \left(\xi-\frac16\right)^2\left(-\frac{ 72 H^2 m^2}{ 6\xi -1} - 216 H^2 \dot{H} + 
 36 \dot{H}^2  - 72 H \ddot{H}\right) }{16 k^3}\right)\right\} \\
-\left(\xi-\frac16\right)\frac{72 H^4  + 72 H^2 \dot{H}  + 
 18 \dot{H}^2   - 216 H^2 \dot{H} (\xi -\frac16)- 
 108 \dot{H}^2 (\xi -\frac16)}{96 \pi^2}-\frac{1-4 \log 2}{128 \pi^2}m^4
 -\frac{H^2 m^2}{96 \pi^2}\notag
\end{gather}
\begin{equation}\label{eq_rhofinalth}\rho^m_\text{gth} = \frac{1}{2\pi^3}\int\limits_0^\infty dk k^2\frac{1}{a^4}\frac{1}{e^{\beta k_0}-1}\left(|\chi_k^\prime|^2+\left(\xi -\frac16\right)2a\Re{\chi_k^\prime\overline{\chi_k}}+\left(k^2+m^2a^2-\left(\xi -\frac16\right)H^2a^2\right)|\chi_k|^2\right)
\end{equation}

In the conformally coupled case $\xi=\frac16$ one can show by a straightforward computation that

\begin{align}\label{eq_rhofinalconformal}
\rho^m_\text{gvac}& = \frac{1}{2\pi^3}\int\limits_0^\infty dk k^2\left\{\frac{1}{2a^4}\left(|\chi_k^\prime|^2+\left(k^2+m^2a^2\right)|\chi_k|^2\right)\right.\notag\\&\quad\left.-\left(\frac{k}{2 a^4}+\frac{m^2}{4 a^2 k}-\Theta(k-ma)\frac{m^4}{16 k^3}\right)\right\} -\frac{1-4 \log 2}{128 \pi^2}m^4
 -\frac{H^2 m^2}{96 \pi^2}\\
 &=\frac{4\pi}{(2\pi)^3}\int\limits_0^\infty dk k^2\frac{1}{2a^4}\left\{\left(|\chi_k^\prime|^2+\left(k^2+m^2a^2\right)|\chi_k|^2\right)-
 \left(|\chi_{0,k}^\prime|^2+\left(k^2+m^2a^2\right)|\chi_{0,k}|^2\right)\right\}\,,\notag
\end{align}
\end{widetext}
\noindent where $\chi_{0,k}$ are the adiabatic modes of order 0, cf. the previous section. This implies that the so-called {\it Hadamard point-splitting regularisation} of the energy density coincides with the so-called {\it adiabatic regularisation of order zero} up to the trace anomaly term and terms which can be subsumed under the regularisation freedom. In the following we analyse the individual state-dependent terms in the energy density.

\subsubsection{Computation of $\rho^m_\text{gvac}$}

\begin{figure}
\includegraphics[width=1\columnwidth]{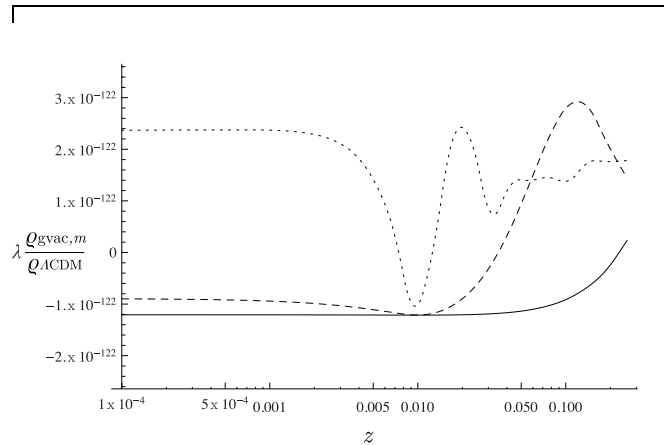}
\caption{\label{fig_rhonormsmallz}$\lambda \rho^m_\text{gvac}/\rho_\text{$\Lambda$CDM}$ for $z<1$ for various values of $m$ (rescaled for ease of presentation). The dotted line corresponds to $m=100H_0$ and $\lambda=10^{-2}$, the dashed line to $m=10H_0$ and $\lambda=1$ and the solid line to $m=H_0$ and $\lambda=10^{2}$. One sees nicely how the energy density is minimal in the support of the sampling function at around $z=10^{-2}$.}
\end{figure}

\begin{figure}
\includegraphics[width=1\columnwidth]{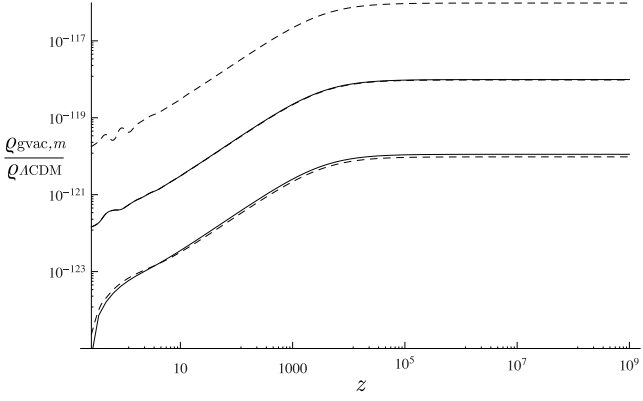}
\caption{\label{fig_rhonormlargez}$\rho^m_\text{gvac}/\rho_\text{$\Lambda$CDM}$ for $z>1$ for various values of $m$. The upper line corresponds to $m=100H_0$, the middle lines to $m=10H_0$ and the lower lines to $m=H_0$; solid lines (dashed lines) indicate results obtained with exact modes (zeroth order adiabatic modes). $\rho^m_\text{gvac}/\rho_\text{$\Lambda$CDM}$ becomes constant for large $z$ because there both energy densities scale like $a^{-4}$, c.f. \eqref{eq_scaling} and the related discussion.}
\end{figure}

Following our general strategy in this section, we first aim to show that in states of low energy on the Robertson-Walker spacetime specified by \eqref{eq_friedmannlcdm} defined by a sampling function of sufficiently large support in time, $\rho^m_\text{gvac}$ is for all $z\in[0,10^9]$ negligible in comparison to the total energy density in the $\Lambda$CDM-model. Results in this direction have been reported in \cite{Degner} for the simplified situation of a de Sitter spacetime background  (corresponding to $\Omega_m=\Omega_r=0$), here we generalise these results to $\Lambda$CDM-backgrounds. One can easily see that $\rho^m_\text{gvac}=0$ in the case of $m=0$. For masses in the range of the Hubble constant $m\simeq H_0$ and states of low energy we have performed numerical computations and found $\rho^m_\text{gvac}/\rho_\text{$\Lambda$CDM} < 10^{-116}$, see Figures \ref{fig_rhonormsmallz}, \ref{fig_rhonormlargez}. To achieve this result, we have rewritten all expressions in terms of the redshift $z$ as a time variable and solved the equation \eqref{eq_modesode} with initial conditions at $z=0$ given by the value and derivative of the adiabatic modes of order zero $\chi_{k,0}$ there. Note that a state of low energy does not depend on the choice of a mode basis, but the choice we made seemed to be numerically favoured. To fix the state of low energy, we chose a sampling function which was a symmetric bump function in $z$ supported in the interval $z\in (10^{-2}, 10^{-2}+10^{-4})$ for definiteness. In order to make the numerical computations feasible, we chose a logarithmic sampling of $k$ with $10^3$ sampling points, where the boundaries of the sampling region have been chosen such that the integrand of $\rho^m_\text{gvac}$, cf. \eqref{eq_rhofinalconformal}, was vanishing in $k$-space to a large numerical accuracy outside of the sampling region for all $z\in[0,10^9]$. We have computed the mode coefficients $c_i(k)$ in the mode basis chosen at each sampling point by a numerical integration in $z$ and finally the energy density by means of a sum over the sampling points in $k$-space. Thus we have approximated the integral in \eqref{eq_rhofinalconformal} by a Riemann sum with logarithmic sampling. As our main aim here is to demonstrate that $\rho^m_\text{gvac}/\rho_\text{$\Lambda$CDM}\ll 1$ in general, we have not performed an extensive analysis of the dependence of $\rho^m_\text{gvac}$ on the width of the sampling function, but we have observed that the maximum amplitude of $\rho^m_\text{gvac}/\rho_\text{$\Lambda$CDM}$ seems to be monotonically growing with shrinking width of the sampling function, in accordance with the computations of \cite{Degner} in deSitter spacetime.

Unfortunately, we have not been been able to compute $\rho^m_\text{gvac}/\rho_\text{$\Lambda$CDM}$ for $m>10^2 H_0$ in the way outlined above because for large masses the modes oscillate heavily, and thus it costs a lot of computer power to solve the mode equation for such a large $z$-interval we are interested in and to the numerical accuracy which is necessary to obtain reliable results for the coefficients of the state of low energy and $\rho^m_\text{gvac}/\rho_\text{$\Lambda$CDM}$. However, realistic field masses in the GeV regime are rather of the order of $10^{42} H_0$. In the numerical computations outlined above we have observed that $\rho^m_\text{gvac}/\rho_\text{$\Lambda$CDM}$ seemed to grow quadratically with $m$, see Figure \ref{fig_rhonormlargez}, but looking at the results of \cite{Degner} in de Sitter spacetime one could maybe expect that $\rho^m_\text{gvac}/\rho_\text{$\Lambda$CDM}$ decreases for large masses. Moreover, even if a potential quadratic growth of $\rho^m_\text{gvac}/\rho_\text{$\Lambda$CDM}$ with $m$ would still imply $\rho^m_\text{gvac}/\rho_\text{$\Lambda$CDM}\ll 1$ for realistic masses and given $\rho^m_\text{gvac}/\rho_\text{$\Lambda$CDM} \sim 10^{-120}$ for $m= H_0$, it would be better to have a more firm understanding of the large mass regime.

In view of the numerical problems for large masses we had to resort to an approximation in order to be able to compute $\rho^m_\text{gvac}/\rho_\text{$\Lambda$CDM}$. In fact, we have taken the adiabatic modes of order zero as basis modes for computing the state of low energy. Of course these modes are not exact solutions of the mode equations, but the failure of these modes to satisfy the exact mode equation is decreasing with increasing mass and thus one can expect that the error in all quantities derived from these modes rather than exact modes is also decreasing with increasing mass. We have checked numerically that the energy density computed with adiabatic modes rather than exact modes matched the 'exact' result quite well already for masses in the regime $m\simeq H_0$, see Figures \ref{fig_rhonormlargez}, \ref{fig_rhonormsmallz}. For more details regarding error estimates for adiabatic modes we refer the reader to \cite{Olver}.

Inserting the adiabatic modes $\chi_{0,k}$ we obtain the following expressions for the coefficients $c_i(k)$ of the states of low energy.

$$c_1(k) = \int dz \,f(z) \left\{\frac{m^4 H}{16 \omega_0(k)^5(1+z)} +\frac{\omega_0(k)(1+z)^3}{2 H}\right\}$$
$$c_2(k) = \int dz \,f(z) \left\{\frac{m^4 H}{16 \omega_0(k)^5(1+z)}-i\frac{m^2}{4\omega_0(k)^2}\right)\times$$
$$\times \exp\left({-2i\int^z_{z_0}\frac{\omega_0(k)}{H}dz^\prime}\right\}$$

We now perform another approximation. We take as a sampling function a Gaussian with mean $z_0$ and variance $\sigma\ll 1$
$$f(z)=\frac{1}{\sqrt{2\pi \sigma^2}}\exp\left(-\frac{(z-z_0)^2}{2\sigma^2}\right)$$
and take the zeroth term of the Taylor expansion of both the expressions in the curly brackets in the integrands of $c_i(k)$ and of the integrand appearing in the exponent of the exponential in $c_2(k)$. Without performing a detailed error analysis we note that this is justified for $\sigma\ll 1$ because the higher coefficients of the associated Taylor series differ from the lowest coefficient roughly by factors of either $\partial_zH/H|_{z=z_0}$ or $H(z_0)/m$, both of which are either smaller than or of order one under the assumption of large masses and a $\Lambda$CDM-background. We can now perform the $z$-integrals, which corresponds to considering the Fourier-transform of $f$ in the case of $c_2$. Using $H_0/m\ll 1$ (and thus $H(z_0)/m\ll1$), we can estimate the resulting coefficients (very roughly) as follows
$$c_1(k) > 1$$
$$|c_2(k)|< \exp\left(-\frac{k^2 \sigma^2}{H(z_0)^2}\right)\exp\left(-\frac{m^2\sigma^2}{H(z_0)^2(1+z_0)^2}\right)\,.$$
For $H(z_0)(1+z_0)/(m\sigma)\ll1$, $|c_2(k)|\ll 1$ and we can approximate the Bogoliubov coefficients $\lambda(k)$ and $\mu(k)$ as $\mu(k) \simeq \frac{|c_2(k)|}{2 c_1(k)}$, $\lambda(k)\simeq 1$ and thus estimate $\rho^m_\text{gvac}$ as

$$|\rho^m_\text{gvac}|<\frac{1}{4 a^4}\int\limits^\infty_0 dk k^2(\mu^2+\mu |\lambda|) \left(|\chi_{0,k}^\prime|^2+\omega^2_0|\chi_{0,k}|^2\right)$$
$$<\frac{1}{ a^4}\int\limits^\infty_0 dk k^2\mu |\lambda| \omega_0< \frac{1}{ a^4}\frac{H(z_0)^3 m}{\sigma^3}\exp\left(-\frac{m^2\sigma^2}{H(z_0)^2(1+z_0)^2}\right)$$
such that, barring our approximations, we indeed get a result which shows that the energy density decreases -- exponentially -- for large masses. Note that for not too small $\sigma$ the bound we found is in general small compared to $\rho_\text{$\Lambda$CDM}$ even if we forget about the exponential because $H_0 m$ is much smaller than the the square of the Planck mass, i.e. $1/G$. We also see that the bound grows with growing $z_0$, i.e. if we 'prepare' the state of low energy further in the past, and that it diverges if the width of the sampling function goes to zero; this is in accord with the results of \cite{Degner} in deSitter spacetime. Note that we could have chosen any rapidly decreasing or even compactly supported sampling function in order to obtain a bound which is rapidly decreasing in $m/H_0$ thus one could say that the result does not depend on the shape of the sampling function as long as its width is not too small. Finally, one could of course directly take the point of view that for large masses the adiabatic modes $\chi_{0,k}$ define 'good states' themselves and conclude that in these states $\rho^m_\text{gvac}=0$ on account of \eqref{eq_rhofinalconformal}.

\subsubsection{Computation of $\rho_\text{gtherm, m}$}

We now proceed to analyse the thermal parts of the state-dependent contributions to the total energy density. Inserting $\xi=\frac16$ in \eqref{eq_rhofinalth}, we find
\begin{equation}\label{eq_thermaldensity}\rho^m_\text{gth} = \frac{1}{2\pi^2}\frac{1}{a^4}\int\limits_0^\infty dk k^2\frac{1}{e^{\beta k_0}-1}\times\end{equation}
$$\times\left(|\chi_k^\prime|^2+\left(k^2+m^2a^2\right)|\chi_k|^2\right)$$
with $k_0=\sqrt{k^2+a_F^2m^2}$. 

Before performing actual computations, we would like to mention a general result about the scaling behaviour of the energy density w.r.t. $a$ \cite{Pinamonti}. To wit, using the equation of motion \eqref{eq_modesode} and the assumption that $H>0$ one can compute the derivative of $Q_k:=|\chi_k^\prime|^2+\left(k^2+m^2a^2\right)|\chi_k|^2$ with respect to $a$ and obtain the following inequalities
\begin{equation}\label{eq_scaling}\frac{k^2+a^2m^2}{k^2+m^2}\frac{Q_k(a=1)}{a^4}\leq\frac{Q_k(a)}{a^4}\leq \frac{Q_k(a=1)}{a^4}\,.\end{equation}
From these one can already deduce that $\rho^m_\text{gth}$ has a scaling behaviour w.r.t. $a$ which lies between $a^{-2}$ and $a^{-4}$ and approaches $a^{-4}$ in the limit of vanishing $a$, in fact this still holds if we replace the Bose-Einstein factors in the generalised thermal states by arbitrary functions of $k$. Moreover \eqref{eq_scaling} also implies that $\rho^m_\text{gvac}$ can not scale with a power of $a$ lower than $-4$ for small $a$ on $\Lambda$CDM backgrounds, c.f. \eqref{eq_rhofinalconformal}.

Proceeding with actual computations we find that in the massless case $\rho^m_\text{gth}$ can be computed exactly and analytically and the result is
\begin{equation}\label{eq_thermaldensitymassless}
\rho^0_{\text{gth}}=\frac{\pi^2}{30}\frac{1}{\beta^4 a^4}\,.
\end{equation}
As in the massless case the state of low energy is the conformal vacuum and the associated generalised thermal state is the conformal temperature state with temperature parameter $\beta=1/T$, this result in fact holds for fields of all spin, i.e. the generalised thermal energy density in this case is always the one in Minkowski spacetime rescaled by $a^{-4}$. Thus a computation with e.g. photons or massless neutrinos yields the same result \eqref{eq_thermaldensitymassless} up to numerical factors due to the number of degrees of freedom and the difference between Bosons and Fermions.

In the massive case it is not possible to compute $\rho^m_\text{gth}$ analytically and exactly, but we have to resort to approximations once more. We recall that the massive scalar field in our model should represent baryonic matter and Dark Matter in a simplified way. Thus we take typical values of $\beta$, $a_F$ and $m$ from Chapter 5.2 in \cite{Kolb:1990vq} computed by means of effective Boltzmann equations. A popular candidate for Dark Matter is a weakly interacting massive particle (WIMP), e.g. a heavy neutrino, for which  \cite{Kolb:1990vq} computes
\begin{equation}\label{eq_kolbvals}x_f = \beta a_F m \simeq 15 + 3\log(m/\text{GeV})\,,\end{equation}
$$a_F \simeq 10^{-12}(m/\text{GeV})^{-1}\,.$$
We shall take these numbers as sample values although working with a scalar field, because for large masses $m\gg H_0$, the ``thermal energy densities'' $\rho^m_\text{gth}$ in generalised thermal states for free fields of spin $0$ and $\frac12$ can be shown to approximately coincide up to constant numerical factors on the basis of the results of \cite{Dappiaggi:2010gt} and \cite[Section IV.5]{Hack:2010iw}.

Considering $m>1$GeV, we can compute $\rho^m_\text{gth}$ approximatively as follows. We recall from the computation of $\rho^m_\text{gvac}$ that for large masses $m\gg H_0$ one can consider the adiabatic modes of order zero $\chi_{0,k}$ as approximative basis modes for the computation of the state of low energy and that with respect to this basis one finds for the coefficients of the state of low energy $\lambda\simeq 1$, $\mu\simeq 0$, thus we can insert those modes in \eqref{eq_thermaldensitymassless} instead of the modes of the state of low energy. Using $m\gg H_0$ once more, we have $|\chi_{0,k}^\prime|^2+\left(k^2+m^2a^2\right)|\chi_{0,k}|^2\simeq \sqrt{k^2+m^2a^2}$ and using $x_f>15$ we can approximate the Bose-Einstein factor in\eqref{eq_thermaldensitymassless} as $1/(e^{\beta k_0}-1)\simeq e^{-\beta k_0}$. Finally we can rewrite the integral in \eqref{eq_thermaldensitymassless} in terms of the variable $y=k/(a_F m)$ and compute, using $a/a_F\gg1$ for the redshift interval $z\in[0,10^9]$ we are interested in,
$$\rho^m_\text{gth}\simeq \frac{1}{2\pi^2}\frac{a_F^3 m^4}{a^3}\int\limits^\infty_0 dy y^2 e^{-x_f\sqrt{y^2+1}}\,.$$
This already gives the desired result $\rho^m_\text{gth}\propto a^{-3}$. The remaining integral can be computed numerically, however, for $x_f \gg 1$ only $y\ll 1$ contribute to the integral and one can approximate $\sqrt{y^2+1}\simeq 1+y^2/2$ and compute 
$$\rho^m_\text{gth}\simeq \frac{1}{(2\pi)^{3/2}}\frac{m}{\beta^3 a^3}x^{\frac32}_f e^{-x_f}\,,$$
which for $a=a_F=1$ (unsurprisingly) coincides with the thermal energy density for massive scalar fields in Minkowski spacetime.

\subsubsection{The total energy density}

Collecting the results of this section, we find for the total energy density of our model

$$\frac{\rho^0 + \rho^m}{\rho_0}\simeq\Omega_\Lambda +  \frac{1}{(2\pi)^{3/2}}\frac{m}{\beta_1^3 a^3\rho_0}x^{\frac32}_f e^{-x_f}\frac{1}{a^3} + \frac{\pi^2}{30\beta_2^4\rho_0}\frac{1}{a^4}\,,$$
where we wrote $\beta_1$, $\beta_2$ in order to emphasise that the generalised thermal states for the massive and massless conformally coupled scalar fields can have different temperature parameters $\beta$. We recall that the thermal contribution of the massless scalar field has been computed exactly, while the one of the massive scalar field is an approximative result. The above result shows that we indeed succeeded in modelling radiation by a massless scalar field and matter/dust by a massive scalar field in suitable generalised thermal states. Obviously, we can choose the free parameters $m$, $\beta_i$, $x_f$ in such a way that the prefactors of the matter and radiation terms have their correct $\Lambda$CDM-values $\Omega_m$ and $\Omega_r$, e.g. for the former we could choose the sample values \eqref{eq_kolbvals} with $m\simeq 100$GeV, and for the latter $1/\beta\simeq 1$K, i.e. the temperature of the CMB. Finally, we model Dark Energy simply by a cosmological constant, which in our context appears as a parametrisation of the freedom in defining energy density as an observable.

Since our description of the the standard cosmological model within quantum field theory on curved spacetime reproduces the energy density of the original $\Lambda$CDM-model up to negligible corrections, it can obviously be matched to the observational data as good as this model.

\section{A natural explanation for Dark Radiation}
\label{sec_dark}

Our analysis in the previous section implies that there exist quantum states in which the total energy density in quantum field theory on curved spacetimes differs from the one in the $\Lambda$CDM-model only by the higher derivative term $\epsilon J_{00}$ and terms which are generally negligible or become important only at redshifts $z\gg 10^9$. The prefactor $\epsilon$ of $J_{00}$ is not determined by the theory but a free parameter so far and it seems advisable to study its impact on the cosmological expansion. Indeed, as our second main result we demonstrate in this section that such term can provide a natural explanation of Dark Radiation. 

To start with, we briefly review the notion of Dark Radiation and the related observations.
The fraction $\Omega_r$ of the radiation energy density in the $\Lambda$CDM-model is computed as
\begin{equation}\label{eq_radiation}\Omega_r=\Omega_\gamma\left(1+\frac78 \left(\frac{4}{11}\right)^{4/3} N_\text{eff}\right)\end{equation}
where $\Omega_\gamma\simeq 5\times 10^{-5}$ is the fraction due to electromagnetic radiation, which can be computed by inserting into
\eqref{eq_thermaldensitymassless} the CMB temperature $T_\text{CMB}\simeq 2.725$K, dividing by today's energy density $\rho_0=3H^2_0/(8\pi G)\simeq 1.33 \times 10^{-11}$eV (and multiplying by two for the two degrees of freedom of the photon). Moreover, $N_\text{eff}$ is the number of neutrino families and the factor $7/8 (4/11)^{4/3}=0.2271$ takes into account that neutrinos are Fermions and moreover ``colder'' than the CMB photons, because they have decoupled from the hot early bath in the universe earlier than electrons and positrons and have thus not been ``heated up'' by the decoupling of the latter like the photons. The standard value for $N_\text{eff}$ is not 3 as one would expect, but rather $N_\text{eff}=3.046$ because the value $7/8 (4/11)^{4/3}$ in \eqref{eq_radiation} is computed assuming e.g. instantaneous decoupling of the neutrinos and corrections have to be taken into account in a more detailed analysis \cite{Mangano:2005cc}; it is customary to take these corrections into account by considering $N=3.046$ as the standard value of the `neutrino family number' rather than changing the factor $7/8 (4/11)^{4/3}$ in this formula, hence the nomenclature $N_\text{eff}$. Consequently, it is convenient to parametrise any contribution to $\Omega_r$ which is not due to electromagnetic radiation and the three neutrino families in the standard model of particle physics by $\Delta N_\text{eff}:=N_\text{eff}-3.046$.

One of the two main observational inputs to determine $\Omega_r$ and thus $N_\text{eff}$ is the primordial fraction of light elements in the early universe as resulting from the so-called Big Bang Nucleosynthesis (BBN), which has occurred at around $z\simeq 10^9$ and thus in the radiation--dominated era, because the nucleosynthesis processes which happened at that time depend sensitively on the expansion rate $H\simeq H_0 \sqrt{\Omega_r}/a^2$, see e.g. \cite{Kolb:1990vq, Kneller:2004jz, Ellis}. The other main observational source for the determination of $N_\text{eff}$ is the cosmic microwave background radiation (CMB). This radiation was emitted at about $z\simeq 1100$, but the CMB power spectrum is sensitive to the expansion before this point, e.g. to the redshift $z_\text{eq}$ at which the energy densities of matter and radiation were equal, see \cite[Section 6.3]{Ade:2013zuv} and the references therein for details; for standard values, $z_\text{eq}\simeq 3000$. 

The observations to date do not give a conclusive value for $N_\text{eff}$, and the value inferred from observational data depends on the data sets chosen. The Planck collaboration \cite{{Ade:2013zuv}} reports e.g. values of $N_\text{eff}=3.36^{+0.68}_{-0.64}$ at 95\% confidence level from combined CMB power spectrum data sets, $N_\text{eff}=3.52^{+0.48}_{-0.45}$ at 95\% confidence level from combining these data sets with direct measurements of the Hubble constant $H_0$ and of the power spectrum of the three-dimensional distribution of galaxies (so-called baryon acoustic oscillation, BAO), and $N_\text{eff}=3.41\pm 0.30$ at 68\% confidence level from combining the CMB power spectrum data sets with BBN data. Yet, one can infer from these values that there is a mild, but not very significant, preference for $\Delta N_\text{eff}>0$. Thus there has been an increasing interest in models which can explain a potential excess in radiation and thus $\Delta N_\text{eff}$, see for instance the recent surveys \cite{DiBari:2013dna, DiValentino:2013qma, Kelso:2013paa, Menestrina:2011mz} and references therein. Most of there models assume additional particles/fields, e.g. a fourth, sterile, neutrino, whereas other consider geometric effects from e.g. modifications of General Relativity. Moreover, in most models $\Delta N_\text{eff}$ is constant and thus affects BBN and CMB physics alike, while in others, e.g. \cite{Fischler:2010xz, Hasenkamp:2011em}, $\Delta N_\text{eff}$ is generated only after BBN and thus affects only CMB physics. 

In the following we shall propose a new and alternative explanation for Dark Radiation which follows naturally from our analysis of the $\Lambda$CDM-model in quantum field theory on curved spacetimes and has the interesting characteristic that it generates a value of $\Delta N_\text{eff}$ which {\it increases} with $z$ and thus affects BBN physics more than CMB physics. To our knowledge, this is the first explanation for Dark Radiation proposed which has this characteristic feature.

\begin{figure}
\includegraphics[width=1\columnwidth]{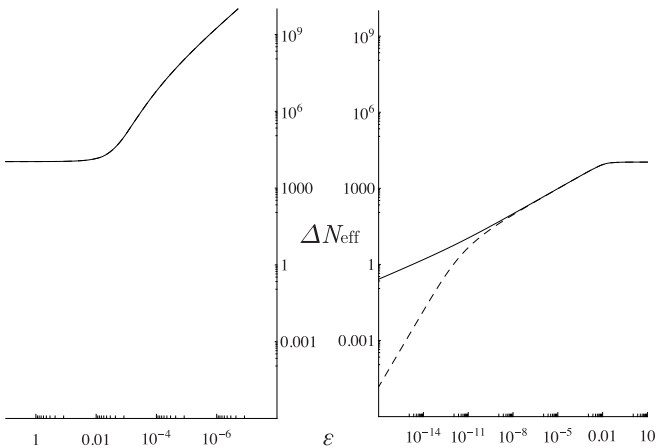}
\caption{\label{fig_deltaneffofalpha}$\Delta N_\text{eff}(z)$ depending on $\epsilon$ for $z=10^9$ (BBN, solid line) and $z=3\times10^3$ (CMB, dashed line). For $\epsilon<0$ and $\epsilon$ positive and large enough, the values at the two redshifts coincide because the maximum value of $\Delta N_\text{eff}(z)$ is reached already for $z<3\times10^3$ in these cases.}
\end{figure}

Following the motivation outlined at the beginning of this section, we solve the equation
\begin{equation}
\label{eq_odebox}
\frac{H^2}{H^2_0}=\Omega_\Lambda + \frac{\Omega_m}{a^3} + \frac{\Omega_r}{a^4} + \epsilon \frac{J_{00}}{H^4_0}\,, 
\end{equation}
which can be rewritten as a second order ordinary differential equation for $H$ in $z$, numerically 
with $\Lambda$CDM-initial conditions $H(z=0)=H_0$, $\partial_zH(z=0)=H_0(3\Omega_m+4\Omega_r)/2$. As before, we consider for definiteness $\Omega_m=0.30$, $\Omega_\Lambda=1-\Omega_m-\Omega_r$, because the exact values of these parameters are not essential for our analysis. Looking at the characteristics of the solution to this ordinary differential equation, it turns out that a non-zero $\epsilon$ generates a time-varying $\Delta N_\text{eff}>0$. In more detail, we define for the solution $H$ of \eqref{eq_odebox}
$$\Delta N_\text{eff}(z):= \frac{\frac{H^2}{H^2_0}-\Omega_\Lambda - \Omega_m(1+z)^3-\Omega_r(1+z)^4}{0.2271(1+z)^4}\,,$$
and sample this observable at the redshift $z=10^9$ associated to BBN physics and at the redshift $z=3000$ associated to CMB physics. We collect our results in Figure \ref{fig_deltaneffofalpha}.

\begin{figure}
\includegraphics[width=1\columnwidth]{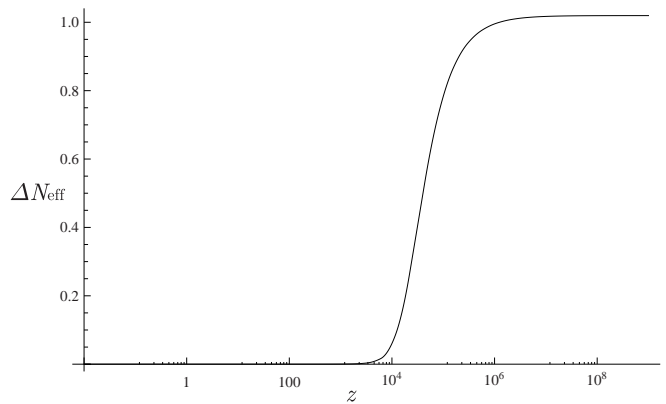}
\caption{\label{fig_deltaneffofz}$\Delta N_\text{eff}(z)$ for $\epsilon=2\times 10^{-15}$.}
\end{figure}

As can be inferred from this figure, $\Delta N_\text{eff}(z)$ is monotonically increasing in $\epsilon$, where positive and negative values of $\epsilon$ result in very different behaviours. For positive values of $\epsilon$ one finds that $\Delta N_\text{eff}(z)$ vanishes in the limit of vanishing $\epsilon$, as one would expect. On the other hand, it turns out that for negative values of $\epsilon$, $\Delta N_\text{eff}(z)$ diverges as $\epsilon$ approaches zero. While this seems to be puzzling at first sight, it fits well with previous qualitative analyses of the effect of the higher derivative term $J_{00}$. In fact, it is known that the inclusion of this higher derivative term can lead to unstable solutions of the semiclassical Einstein equations, where for $\epsilon<0$ ($\epsilon>0$) the class of solutions we consider here, effectively fixed by the $\Lambda$CDM initial conditions, turns out to be unstable (stable), see e.g. \cite{Anderson1, FW96, Haensel, Koksma, ParkerSimon, Star}. Thus, the divergence of $\Delta N_\text{eff}(z)$ as $\epsilon$ approaches zero from below can be just interpreted as a sign of this instability. 

In \cite{Wald4, DFP}, $\epsilon=0$ has been chosen on conceptual grounds in order to discard unstable solutions altogether. However, as we see here a non-zero $\epsilon$ can have interesting phenomenological implications. After all, taking quantum field theory on curved spacetimes seriously, $\epsilon$ is a free parameter of the  theory, which we can only fix in a more fundamental theory or by observations. Indeed, we see  in Figure \ref{fig_deltaneffofalpha} that $\epsilon<0$, corresponding to an unstable solution of the semiclassical Einstein equation, is already ruled out by observations because it generally leads to $\Delta N_\text{eff}(z)\gg 1$ which is  certainly not compatible with value of $\Delta N_\text{eff}\simeq 0.5-1.0$ inferred from observations as mentioned above. On the other hand we see that in order to not exceed $\Delta N_\text{eff}=1$ at both BBN and CMB we have to choose $0\le \epsilon < 2\times 10^{-15}$, thus, without performing a detailed fit of BBN and CMB data, we can say that the values for $\Delta N_\text{eff}$ reported e.g. by the Planck collaboration in \cite{{Ade:2013zuv}} give an upper bound of about $2\times 10^{-15}$ for $\epsilon$. We plot $\Delta N_\text{eff}(z)$ for redshifts $0<z<10^9$ in Figure \ref{fig_deltaneffofz}. As already anticipated in Figure \ref{fig_deltaneffofalpha}, one can nicely see how $\Delta N_\text{eff}(z)$ is monotonically growing in $z$, with $\Delta N_\text{eff}(z=0)=0$ as fixed by our initial conditions. Moreover one can see clearly that if one wants to meet the bounds on $\Delta N_\text{eff}$ at the BBN redshift, the excess in the effective number of neutrinos at the CMB is negligible, which is the characteristic signature of this potential explanation for Dark Radiation. We have not considered the influence of the initial conditions for \eqref{eq_odebox} on $\Delta N_\text{eff}$, but we expect that for the initial conditions compatible with low-$z$ observational data such as supernova type Ia data and baryon acoustic oscillation data, $\Delta N_\text{eff}$ will not differ considerably from the form we found as these data will not allow for large deviations in the initial conditions from the $\Lambda$CDM ones we chose.

As a further, rather pedagogical remark, we would like to comment on the fact that, for large absolute values of $\epsilon$, $\Delta N_\text{eff}$ does not depend on the sign of $\epsilon$, as can be seen from Figure. \ref{fig_deltaneffofalpha}. This phenomenon can be understood as follows. Naturally, for large absolute values of $\epsilon$, the other terms in \eqref{eq_odebox} become negligible and one effectively solves for $J_{00}=0$. The solution of this ordinary differential equation with initial conditions $H(z=0)=c$, $\partial_zH(z=0)=d$ is
$$H=\frac{c^{1/3}d^{2/3}\left(\frac{2c}{d}-1+(1+z)^3\right)^{2/3}}{2^{2/3}}$$
and thus, inserting the $\Lambda$CDM initial conditions $H(z=0)=H_0$, $\partial_zH(z=0)=H_0(3\Omega_m+4\Omega_r)/2$ we find for large $z$  $N_\text{eff}(z)\simeq 10^4$ as in Figure \ref{fig_deltaneffofalpha}.

While our analysis is to our knowledge the first attempt to bound $\epsilon$ with cosmological observations, it is not the first attempt to determine it with observations at all. In fact, the effect of higher-derivative corrections to General Relativity has already been analysed in the past, and since the tensors $I_{\mu\nu}$ and $J_{\mu\nu}$ in \eqref{eq_nonunique} can be obtained as variational derivatives with respect to the metric of the Lagrangeans $\sqrt{-g}R^2$ and $\sqrt{-g}R_{\mu\nu}R^{\mu\nu}$, they have been considered in these analyses as well. To wit, the Lagrangean
$$L=\sqrt{-g}\left(\frac{R}{16\pi G}+c_1 R^2 + c_2 R_{\mu\nu}R^{\mu\nu}\right)$$
leads to the Newtonian potential of a point mass $m$ \cite{Stelle:1977ry}
\begin{equation}\label{eq_newton}
\phi = \frac{-m G}{r}\left(1+\frac{1}{3}e^{-m_1 r}-\frac{4}{3}e^{-m_2 r}\right)\end{equation}
$$m_1=\frac{1}{\sqrt{32 \pi G(-3c_1-c_2)}}\qquad m_2=\frac{1}{\sqrt{16\pi G c_2}}\,.$$
Using recent data \cite{Kapner:2006si} from torsion-balance experiments to test the gravitational inverse-square law at $\sim 10^{-4}$m and assuming that the two Yukawa corrections don't cancel each other at this length scale, one obtains $-c_1,c_2<10^{61}$ \cite{Calmet:2008tn}. To compare this with our results, we recall that in our treatment these higher curvature terms appear on the right hand side of the semiclassical Einstein equation and that we have computed in units of $H_0$, thus we have 
$$\epsilon=\frac{(-3c_1-c_2)8\pi G H^2_0}{3}\simeq (-3c_1-c_2)10^{-121}$$
which would imply $\epsilon < 10^{-60}$ and thus a stronger bound then the one we inferred from cosmological observations. Of course such a low value of $\epsilon$ leads to $\Delta N_\text{eff}\ll 1$ at both BBN and CMB and thus no Dark Radiation would be generated.

Notwithstanding, there are still several aspects of our analysis which are of interest. First of all, our bound on $\epsilon$ is completely independent from the one inferred from laboratory experiments and can thus be considered as an additional confirmation of those results. Moreover, it is still possible that the Yukawa corrections in \eqref{eq_newton} cancel each other on the length scales relevant for the experiments described in \cite{Kapner:2006si}, such that $\epsilon$ could be as large as our upper bound, which in this case would give a real bound on one and hence both Yukawa corrections. Finally, the bounds inferred from \cite{Kapner:2006si} and from our analysis stem from phenomena on completely different length scales. As a rough estimate we note that the diameter of our observable universe, which today is about $6/H_0\simeq 10^{27} $m, was at e.g. $z=10^9$ still $10^{18}$m and thus much larger than the submillimeter scales relevant for the torsion-balance experiments. Thus it could be that effects we have not considered so far, e.g. state-dependent effects which are due to the small-scale structure of the quantum states we have fixed only on cosmological scales so far, affect the comparison between the two different sources of input for the determination of $\epsilon$.

\section{Conclusions}
\label{sec_conclusions} 

We have demonstrated that it is possible to understand the cosmological evolution for redshifts $z<10^9$ as described in the $\Lambda$CDM-model entirely in terms of quantum field theory in curved spacetime, by computing the energy density in generalised thermal quantum states and showing that the state and field parameters can be chosen such as to match the energy density in the $\Lambda$CDM-model up to small corrections. 

One of these corrections, quantified by a parameter $\epsilon$, occurred due to higher-derivative terms appearing as renormalisation freedom of the energy density of any quantum state. We have demonstrated that this correction can constitute a natural explanation for Dark Radiation with the characteristic signature of leading to a time-varying effective number of neutrino families $N_\text{eff}$ which decays in time and have obtained the bound $\epsilon<2\times 10^{-15}$ by comparison with experimental data, which is compatible with the $\Lambda$CDM-value $\epsilon=0$. A conservative interpretation of laboratory experiments leads to a stronger bound $\epsilon<\times 10^{-60}$ which cancels any Dark Radiation effects, but we have argued that there are possibilities to evade this stronger bound. Thus we believe that it is worth to include this new parameter in further analyses of the $\Lambda$CDM-model parameters.

An additional correction to the $\Lambda$CDM-model appears due to the so-called trace anomaly. This contribution to the energy density is negligible for redshifts $z<10^9$ but can have considerable impact on the cosmological evolution at larger redshifts as discussed already in the framework of the so-called Starobinski-inflation \cite{Star}. Note that, in contrast to the renormalisation freedom of the energy density quantified by the in principle free model parameter $\epsilon$, the trace anomaly is fixed by the field content and thus {\it predicted} by quantum field theory in curved spacetime, if one accepts the validity of the semiclassical Einstein equations up to the regimes where the trace anomaly becomes important.

Finally, potential further corrections to the $\Lambda$CDM-model can    come from specifics of the quantum state we have neglected in our analysis. We have chosen the quantum states in our discussion such that their characteristic energy density was entirely of thermal nature, but we have seen that also pure, non-thermal states can have  contributions to the energy density which scale like $a^{-4}$, cf. Figure \ref{fig_rhonormlargez}. It could be that there exist states which are compatible with observations and have sizable energy-density contributions of this kind; these states would then provide a further alternative explanation for Dark Radiation which does not call for the introduction of new particles respectively fields.


%





\vskip .2cm

\begin{acknowledgments}
The work of TPH is supported by a research fellowship of the Deutsche Forschungsgemeinschaft (DFG). The author would like to thank Claudio Dappiaggi and Nicola Pinamonti for continuing illuminating discussions on the topic of this paper as well as numerous others. Jan M\"oller deserves sincere thanks for thoroughly suggesting the confrontation of quantum field theory on curved spacetimes with the standard cosmological model.
\end{acknowledgments}


\end{document}